\documentclass{article}
\usepackage{spconf}
\usepackage{amsmath,amssymb,amsfonts,amsthm}
\usepackage{bm, multirow}
\usepackage{algorithmic}
\usepackage{graphicx}
\usepackage{textcomp}

\newcommand{\ignore}[1]{}
\newcommand{\CL}{\mbox{CL}}

\theoremstyle{plain}
\newtheorem{con}{Condition}
\newtheorem{thm}{Theorem}
\newtheorem{lemma}{Lemma}

\title{Extending the Use of MDL for High-Dimensional Problems: Variable Selection, Robust Fitting, and Additive Modeling$^\star$
\thanks{$^\star$A\lowercase{n abridged version of this paper will appear in the} {\it P\lowercase{roceedings of the 47th} IEEE I\lowercase{nternational} C\lowercase{onference on} A\lowercase{coustics}, S\lowercase{peech, and} S\lowercase{ignal} P\lowercase{rocessing} (ICASSP)}, 2022.}}

\name{Zhenyu Wei$^{\dagger}$ \qquad Raymond K. W. Wong$^{\ddagger}$ \qquad Thomas C. M. Lee$^{\dagger}$}
\address{$^{\dagger}$University of California, Davis \qquad $^{\ddagger}$Texas A\&M University}

\begin{document}
\maketitle

\begin{abstract}
In the signal processing and statistics literature, the minimum description length (MDL) principle is a popular tool for choosing model complexity.  Successful examples include signal denoising and variable selection in linear regression, for which the corresponding MDL solutions often enjoy consistent properties and produce very promising empirical results.  This paper demonstrates that MDL can be extended naturally to the high-dimensional setting, where the number of predictors $p$ is larger than the number of observations $n$.  It first considers the case of linear regression, then allows for outliers in the data, and lastly extends to the robust fitting of nonparametric additive models.  Results from numerical experiments are presented to demonstrate the efficiency and effectiveness of the MDL approach.
\end{abstract}
\begin{keywords}
denoising,
heavy-tailed errors, 
outliers, 
spline fitting,
variable screening
\end{keywords}

\section{Introduction}
\label{sec:introduction}
The minimum description length (MDL) principle \cite{rissanen1989stochastic,rissanen2007information} has long been successfully applied to perform signal denoising \cite{cohen1999translation,rissanen2000mdl,roos2009mdl} and model selection in regression and time series problems \cite{Aue-et-al14,cheung2017consistent,gao2017nonparametric,kallummil2016high,Lee00:MDLspline,schmidt2011consistency,wong2010structural}.  This paper extends MDL to solve some high-dimensional problems, including linear regression, nonparametric additive models, and robust fitting.  Notice that this paper does {\bf not} claim that MDL is the only approach for doing so.  Instead, it shows that MDL can be extended to solve these problems in a conceptually clean and natural manner, and produce excellent results.


A typical description of the high-dimensional linear regression problem is as follows.  Let $\bm{y} = (y_1,\ldots, y_n)^T$ be a vector of $n$ responses and $\bm{x}_i$ be a $p$-variate predictor variable for $y_i$.  Write $\bm{X} = (\bm{x}_1,\ldots,\bm{x}_n)^T$ as the design matrix of size $n\times p$.  The observed responses and the predictors are related by the linear model
\begin{equation}
	\bm{y} = \bm{X\beta} + \bm{\varepsilon}, 
	\label{model:1}
\end{equation}
where $\bm{\beta} = (\beta_1,\ldots,\beta_p)^T$ is a vector of unknown parameters and $\bm{\varepsilon} = (\varepsilon_1,\ldots,\varepsilon_n)^T$ is a vector of i.i.d.~random Gaussian errors with zero mean and unknown variance $\sigma^2$.  It is assumed that $p\gg n$, making this high-dimensional regression problem different from the classical multiple regression regression problem for which $p < n$.

When $p\gg n$, one needs to assume that the number of significant predictors in the true model is small; i.e., the true model is sparse.  The problem is then to identify which $\beta_j$'s are non-zero.  This is sometimes known as the variable selection problem, and this paper applies the MDL principle to derive a solution.  Although there are existing methods for addressing this problem \cite{fan2010selective}, this paper seems to be one of the earliest attempts that MDL is being applied and carefully studied in high-dimensional settings.

This paper also considers robust fitting, by relaxing the Gaussianity assumption on $\bm{\varepsilon}$ and allowing the presence of heavy-tailed errors or outliers in the response.  It achieves this goal by modeling the error component with the Laplace distribution. 

Lastly this paper extends the variable selection problem for~(\ref{model:1}) to high-dimensional nonparametric additive models, defined as
\begin{equation}
	y_i = \mu + \sum_{j=1}^p f_j(x_{ij}) + \varepsilon_i, \quad i = 1, \ldots, n,
	\label{model:additive}
\end{equation}
where $\mu$ is an intercept term, the $f_j$'s are unknown nonparametric functions, and $x_{ij}$ is the $j$th covariate of $\bm{x}_i$.  Again, we consider $p \gg n$ and impose the sparsity assumption.  We select the significant predictors, as well as allowing the possibility of outliers.  To the best of our knowledge, this is the first time that outlier-resistant estimation and variable selection for high-dimensional nonparametric additive models is considered.  

Below we first provide some background on MDL.  We then present our new MDL solutions to the problems of high-dimensional linear regression, nonparametric additive models, and robust fitting.  Both theoretical and empirical properties of our proposed methods will also be reported.

\section{A Brief Description of the Minimum Description Length (\text{MDL}) Principle}
\label{sec:bg}
In model selection problems the \text{MDL} principle defines the best fitting model as the one that produces the shortest code length of the data \cite{rissanen1989stochastic, rissanen2007information}.  In this context the code length of an object can be treated as the amount of memory space that is required to store the object.  Of course comparing code lengths is neither the only nor the best approach for defining a best fitting model, but it is still a sensible one.  It is because a common feature of a good encoding (or compression) scheme and a good statistical model is the ability to capture the regularities, or patterns, hidden in the data.

There are different versions of MDL, and this paper focuses on the so-called two-part codes.  When applying this, it is common to split the code length for a set of data into two parts: (i) a fitted model plus (ii) the data ``conditioned on" the fitted model; i.e., the residuals.  If we denote the data as $\bm{y}$, any fitted model as $\hat{\bm{\theta}}$, and the residuals as $\hat{\bm{e}}=\bm{y}-\hat{\bm{y}}$, where $\hat{\bm{y}}$ is the fitted value of $\bm{y}$, we split $\bm{y}$ into $\hat{\bm{\theta}}$ plus $\hat{\bm{e}}$.  Notice that knowing $\hat{\bm{\theta}}$ and $\hat{\bm{e}}$ can completely retrieve $\bm{y}$.

If $\CL(z)$ denotes the code length of an object $z$, we have 
$$\CL(\bm{y}) = \CL(\hat{\bm{\theta}}) + \CL(\hat{\bm{e}}|\hat{\bm{\theta}}).$$
Note that in this expression it is stressed that $\hat{\bm{e}}$ is conditional on $\hat{\bm{\theta}}$; i.e., different $\hat{\bm{\theta}}$'s would give different $\hat{\bm{e}}$'s.  Now the task is to find an expression for $\CL(\bm{y})$ so that the best \text{MDL} $\hat{\bm{\theta}}$ can be defined and obtained as its minimizer.

\section{High-Dimensional Linear Regression}
\label{sec:linear}
We first consider variable selection for model~(\ref{model:1}).  Let $S$ be a subset of $\{1, \ldots, p\}$.  If $j \in S$, it means $\beta_j$ is significant. Hence $S$ can be used to represent any candidate model.  Denote the corresponding design matrix as $\bm{X}_S$, and the maximum likelihood estimate of the corresponding coefficients $\bm{\beta}_S$ as $\hat{\bm{\beta}}_S$.  Also, let $L(\cdot)$ be the likelihood function and $|S|$ be the number of elements in $S$, i.e., the number of significant $\beta_j$'s.  It is shown in Section \ref{derive:1} that a \text{MDL} criterion for the model specified by $S$ is
	\begin{eqnarray}
	\text{MDL}(S) & = & -\log L(\bm{y},\bm{X}_S\hat{\bm{\beta}}_S) + \frac{|S|}{2}\log(n) + |S|\log(p) \nonumber \\
	& = & \frac{n}{2}\log\bigg\{\frac{(\bm{y}-\bm{X}_S\hat{\bm{\beta}}_S)^T(\bm{y}-\bm{X}_S\hat{\bm{\beta}}_S)}{n}\bigg\} \nonumber \\
	& & + \frac{|S|}{2}\log(n) + |S|\log(p) \nonumber \\
	& = &
	\frac{n}{2}\log\left(\frac{\text{RSS}}{n}\right) + \frac{|S|}{2}\log(n) + |S|\log(p),
	\label{eqn:mdllinear}
	\label{EQN:MDLLINEAR}
\end{eqnarray}
where 
$$\mbox{RSS}=(\bm{y}-\bm{X}_S\hat{\bm{\beta}}_S)^T(\bm{y}-\bm{X}_S\hat{\bm{\beta}}_S)$$
is the residual sum of squares.  When comparing to the classical \text{MDL} criterion for $p <n$, this new $\text{MDL}(S)$ has an additional penalty term $|S|\log(p)$, which coincidentally shares the same asymptotic order as the corresponding penalty term in EBIC of \cite{chen2008extended}.  We note that, however, $\text{MDL}(S)$ is different from EBIC for finite samples.

\subsection{Derivation of \text{MDL}}
\label{derive:1}
This section outlines the derivation of~(\ref{eqn:mdllinear}).  Here the parameter vector estimate $\hat{\bm{\theta}}$ is $\hat{\bm{\beta}}_S$, so we begin with 
$$\text{MDL}(S) = \CL(\bm{y}) = \CL(\hat{\bm{\beta}}_S) + \CL(\hat{\bm{e}}|\hat{\bm{\beta}}_S).$$

According to \cite{rissanen1989stochastic}, the code length for encoding an integer $N$ is approximately $\log_2N$ bits. To encode $\hat{\bm{\beta}}_S$, one first needs to identify which of the $|S|$ predictors are selected. Since each of the $|S|$ predictors can be uniquely identified by an index in $\{1, \ldots, p\}$, it takes a total of $|S|\log_2p$ bits to encode this information.  Next, the corresponding parameter estimates need to be encoded.  In \cite{rissanen1989stochastic} it is demonstrated that if a maximum likelihood estimate of a real-valued parameter is computed from $N$ data points, then it can be effectively encoded with $\frac{1}{2}\log_2N$ bits.  This gives the total code length for the $|S|$ parameter estimates as $\frac{|S|}{2}\log_2n$, and hence
\begin{equation}
	\CL(\hat{\bm{\beta}}_S) = |S|\log_2p + \frac{|S|}{2}\log_2n.
	\label{eqn:linearpart1}
\end{equation}
Notice that in classical applications of MDL for problems with $p \ll n$, the term $|S|\log_2p$ is often omitted as it is relatively small compared with $\frac{|S|}{2}\log_2n$.  However, when $p$ is comparable to $n$ or even $p\gg n$ , this term cannot be omitted as otherwise it will give erratic results.

Now it remains to calculate $\CL(\hat{\bm{e}}|\hat{\bm{\beta}}_S)$, and it is shown in \cite{rissanen1989stochastic} that this is equal to the negative of the log of the likelihood of $\hat{\bm{e}}$ conditioned on $\hat{\bm{\beta}}_S$.  For the present problem, it simplifies to 
\begin{equation}
	\CL(\hat{\bm{e}}|\hat{\bm{\beta}}_S)=  \frac{n}{2}\log_2 \left(\frac{\mbox{RSS}}{n}\right).
	\label{eqn:linearpart2}
\end{equation}
Now by changing $\log_2$ to $\log$ and combining~(\ref{eqn:linearpart1}) and~(\ref{eqn:linearpart2}), one obtains $\text{MDL}(S)$ in~(\ref{eqn:mdllinear}).

\subsection{Practical Minimization of (\protect{\ref{EQN:MDLLINEAR}})}
\label{proced:1}
In practice minimizing~(\ref{eqn:mdllinear}) is not a trivial task, especially when $p \gg n$.  This subsection presents a three-stage procedure that aims to locate a good approximated minimizer of~(\ref{eqn:mdllinear}).  The first stage is to apply a screening procedure to remove a large number of insignificant predictors, so that we will only have to consider the remaining $m$ predictors, where $m<n$.

Then in the second stage the lasso \cite{tibshirani1996regression} method is applied to obtain a nested sequence of $m$ candidate models.  Lastly, the $\text{MDL}(S)$ values for these $m$ candidate models are calculated and the one with the smallest value is taken as the final, best fitting model. 

\emph{ Stage 1: Screening.}  The goal here is to remove a lot of non-significant predictors quickly with high confidence.  We propose using the sure independence screening (SIS) procedure of \cite{fan2008sure}.  The idea is to rank the predictors according to the magnitudes of their sample correlations with the response variable, and keep the $m$ largest ones.  More precisely, let $\bm{\omega} = (\omega_1,...,\omega_p)^T = \bm{X}^T\bm{y}$, where we assume that each column of the $n\times p$ design matrix $\bm{X}$ has been standardized with mean zero and variance one.  We keep the $m$ predictors that have the largest $m$ values of $|w_j|$, and collect them in $S^*$; i.e., 
$$S^* = \{1\leq j\leq p:|\omega_j|\text{ is among the first $m$ largest of all}\}.$$
This reduces the number of possible predictors $p \gg n$ to a more manageable number $m$.  This SIS procedure will remove those predictors that have weak marginal correlations with the response, and has been shown to possess excellent theoretical and empirical properties (e.g., see \cite{fan2008sure}).  In practice we set $m=n-1$. 

\emph{ Stage 2: Lasso fitting.}  Lasso was proposed by \cite{tibshirani1996regression} to perform variable selection and shrinkage estimation for linear models.  It produces a so-called solution path from which a sequence of nested models can be obtained.  In the original lasso, cross-validation was suggested to choose a final model from these nested models.  Here, however, we simply apply lasso to $S^*$ and obtain $m$ nested models.  Given the LARS algorithm \cite{efron2004least}, this step can be performed very efficiently.

\emph{ Stage 3: $\text{MDL}(S)$ Calculation.}  Here we use~(\ref{eqn:mdllinear}) to choose a final best fitting model from those nested models obtained above.  However, given the shrinkage nature of lasso, the parameter estimates of these nested models obtained from above are shrunk towards zero.  Therefore, these estimates (and other quantities derived from them such as RSS) should not be used for the calculation of~(\ref{eqn:mdllinear}).  Thus, for each of the nested models, we use maximum likelihood to estimate the unknown parameters, and use these estimates to calculate~(\ref{eqn:mdllinear}).  The model that gives the smallest value of~(\ref{eqn:mdllinear}) is taken as the final model.

\section{Theoretical Properties}
\label{sec:theory}

This section presents some theoretical backup for the above MDL criterion for high-dimensional regression.  Let $S_0$ be the index set of the true model, and $$\bm{\mu}=E(\bm{y})=\bm{X}_{S_0}\bm{\beta}_{S_0}.$$
Define the projection matrix for any $S \subset \{1, \ldots, p\}$ as 
$$\bm{P}_S = \bm{X}_S(\bm{X}^T_S\bm{X}_S)^{-1}\bm{X}^T_S,$$ 
and write  $\delta(S) = \|\bm{\mu}-\bm{P}_S\bm{\mu}\|^2$,
with $\|\cdot\|$ being the Euclidean norm. Clearly, if $S_0\subset S$, we have $\delta(S) = 0$. In our theoretical analysis,  we need the following identifiability condition, which is similar to the condition stated in \cite{chen2008extended}.

\begin{con}
	\label{con:1}
	(Asymptotic identifiability) The true model $S_0$ is asymptotically identifiable if 
	$$\lim\limits_{n\to\infty}\min\left\{\frac{\delta(S)}{\log{n}}: S\ne S_0, |S|\le k|S_0|\right\} = \infty$$
	for some fixed $k>1$.
\end{con} 
Roughly speaking, a true model is asymptotically identifiable if no other model of finite size can predict the response as well as the true model. 
We now have the following theorem.

\begin{thm}
	\label{thm:1}
	\label{THM:1}
	Consider a data set $\{(\bm{x}_i, y_i): i = 1,\ldots, n\}$ from model~(\ref{model:1}).
	Suppose Condition~\ref{con:1} holds
	and $p = O(n^{\gamma})$ for some fixed $\gamma$. Also assume $\varepsilon_1,\dots,\varepsilon_n\overset{\text{i.i.d.}}{\sim} N(0, \sigma^2)$. Then we have
	$$
	P\left(\min_{S\ne S_0, |S|\le k|S_0|}\text{MDL}(S)>\text{MDL}(S_0)\right)\to 1 
	$$
	as $n\to\infty$.
\end{thm}

We note that the penalties in \eqref{eqn:mdllinear}, derived from the MDL framework, share the same order as the one in EBIC.  Indeed, the proof of Theorem~\ref{thm:1} can be constructed by using similar idea in \cite{chen2008extended}, and can be found in the appendix.  Theorem~\ref{thm:1} indicates that, as $n\to\infty$, the probability that $\text{MDL}(S)$ wrongly selects a model of a similar size other than the true model goes to zero.

\section{Robust Fitting for High-Dimensional Linear Regression}
\label{sec:robust}
This section demonstrates how robust estimation for high-dimensional linear regression can be handled by MDL.  We first review some existing work in this area.

A robust version of lasso was proposed by \cite{wang2007robust}.  It replaces the $L_2$ norm for measuring data fidelity with a least absolute deviation type of norm, which will make the method less sensitive to the presence of outliers.  However, this method was not originally designed for the high-dimensional setting.  Another method termed RLARS, short for robust least angle regression, was developed by \cite{khan2007robust}.  It is a robust version of the LARS algorithm \cite{efron2004least}, and it essentially uses a robust correlation to rank and select the most important variables.  More recently, a sparse and regularized version of the least trimmed squares (LTS) was proposed by \cite{alfons2013sparse}, which introduces an $L_1$ penalty to the LTS estimator.  This sparse LTS estimator can also be interpreted as a trimmed version of the lasso.  However, this method is computationally expensive.

Robust fitting can be embedded into the MDL framework.  A natural approach is to adopt a heavy tail distribution for the errors $\varepsilon_i$'s to allow for outliers.  For this we suggest using the zero mean Laplace distribution; i.e., 
$$\varepsilon_i \overset{\text{i.i.d.}}{\sim} \mbox{Laplace}(0,b),$$
where $b$ is a scale parameter.

Similar to Section~\ref{derive:1}, it can be shown that the \text{MDL} criterion for robust fitting with a model specified by $S$ is
\begin{eqnarray}
	\text{MDL}_{\rm robust}(S) & = & -\log L(\bm{y},\bm{X}_S\hat{\bm{\beta}}_S) + \frac{|S|}{2}\log(n)\nonumber \\
	& & + |S|\log(p) \nonumber \\
	& = & n\log\bigg(\frac{\sum_{i=1}^n|y_i-\bm{x}_{S,i}\hat{\bm{\beta}}_S|}{n}\bigg)  \nonumber \\ 
	& &  + \frac{|S|}{2}\log(n) + |S|\log(p) \nonumber \\
	& = &
	n\log\bigg(\frac{\text{SAE}}{n}\bigg) + \frac{|S|}{2}\log(n) \nonumber \\
	& & + |S|\log(p).
	\label{eqn:mdllinearrobust}
\end{eqnarray}
In the above $\text{SAE}=\sum_{i=1}^n|y_i-\bm{x}_{S,i}\hat{\bm{\beta}}_S|$ is the sum of absolute errors, with $\bm{x}_{S,i}$ denoting the $i$-th row of $\bm{X}_S$.  Note that the maximum likelihood estimate for the scale parameter $b$ is $\hat{b} = \text{SAE}/n$.

Practical minimization of~(\ref{eqn:mdllinearrobust}) can be achieved in a similar fashion as the 3-stage procedure described in Section~\ref{proced:1}.  To be more specific, Stage~1 remains the same, while in Stage~2 the robust LARS method of \cite{khan2007robust} is used in place of the original lasso, and in Stage~3 the MDL criterion~(\ref{eqn:mdllinearrobust}) is used instead of~(\ref{eqn:mdllinear}) when calculating the MDL values.

\section{High-Dimensional Nonparametric Additive Models}
\label{sec:additive}
This section extends our work to the high-dimensional nonparametric additive models~(\ref{model:additive}).  The goal is to select those significant ones from the functions $f_1, \ldots, f_p$, as well as to estimate them nonparametrically.  We first discuss the use of splines for modeling the $f_j$'s.

\subsection{Spline Modeling for Additive Functions}
Briefly, a spline function is a piecewise polynomial function.  The locations at which two adjacent pieces join are called knots.  Here we state their standard conditions and definition.

Suppose that $x \in [a,b]$ for some finite numbers $a<b$ and that
$E(y^2)<\infty$.

To ensure identifiability, it is assumed $E\{f_j(x)\}=0$ for $j=1,\ldots, p$. Let $K$ be the number of knots for a partition of $[a,b]$ that satisfy specific conditions stated for example in \cite{lai2012fixed}.  Let $\mathcal{S}_n$ be the collection of functions $s$ with domain $[a,b]$ satisfying the following two conditions: (i) $s$ is a polynomial of degree $l$ (or less) on each subinterval, and (ii) for any two integers $l$ and $l'$ satisfying $l\geq 2$ and $0\leq l'<l-1$, $s$ is $l'$-times continuously differentiable on $[a,b]$.  Then there exists a normalized B-spline basis $\{\varphi_{k}(\cdot), k=1,\ldots,d_n\}$ such that for any $s \in \mathcal{S}_n$, we have

\begin{equation}
	s(x) = \sum_{k=1}^{d_n} \alpha_{k} \varphi_{k}(x),
	\label{eqn:spline}
\end{equation}
where $\alpha_{k}$ is the coefficient of the basis function $\varphi_{k}(x)$ for $k=1,\ldots, d_n$ with $d_n=K+l$.  Since $\mathcal{S}_n$ is a relatively rich class of smooth functions, in this paper, for the reason of speeding up technical calculations, we shall assume that the spline representation~(\ref{eqn:spline}) is exact for the functions $f_j$'s.  In other words, for $j=1, \ldots, p$, it is assumed that
\begin{equation}
	f_j(x) = \sum_{k=1}^{d_n} \alpha_{jk} \varphi_{k}(x),
	\label{eqn:fjspline}
\end{equation}
where $\alpha_{jk}$'s are the corresponding coefficients of the bases $\varphi_{k}(x)$'s.

\subsection{MDL Criteria}
Recall that for the fitting of the high-dimensional nonparametric additive models~(\ref{model:additive}), we aim to select those significant functions from $f_1, \ldots, f_p$, as well as to estimate them nonparametrically.  For any candidate model, denote the number of significant $f_j$'s as $q$, and the number of basis functions used for each $f_j$ as $d_n$.  Using similar steps as in Section~\ref{derive:1}, it can be shown that an MDL criterion for fitting~(\ref{model:additive}) is:
\begin{eqnarray}
	\text{MDL}_{\rm additive}(S)
	& = & \frac{n}{2}\log\bigg(\frac{\text{RSS}}{n}\bigg) + \frac{qd_n}{2}\log(n) \nonumber \\
	& & + q\log(p).
	\label{eqn:mdladditive}
\end{eqnarray}
One can also perform robust fitting as in Section~\ref{sec:robust} above, and the resulting MDL criterion is

\begin{eqnarray}	
	\text{MDL}^{\rm robust}_{\rm additive}(S) & = & n\log\bigg(\frac{\text{SAE}}{n}\bigg) + \frac{qd_n}{2}\log(n) \nonumber \\
	& & + q\log(p).
	\label{eqn:mdladditiverobust}
\end{eqnarray}

\subsection{Practical Minimization}
MDL criteria~(\ref{eqn:mdladditive}) and~(\ref{eqn:mdladditiverobust}) can be minimized in a similar manner as in Section~\ref{proced:1}.  In Stage~1 we screen out most of the non-significant function $f_j$'s.  However, instead of using SIS which was designed for linear regression problem, we use the nonparametric independence screening (NIS) procedure of \cite{fan2011nonparametric}.

In Stage~2 we apply the group lasso of \cite{yuan2006model} to obtain a nested sequence of models.  The reason the original lasso is not applicable here is that, all the coefficients $\alpha_{jk}$'s belonging to the same function $f_j$ should either be kept or removed together (see~(\ref{eqn:fjspline})), and the original lasso will not guarantee this.  On the other hand, the group lasso was designed for this purpose.

In the last stage we first re-fit all the nested models obtained from Stage~2 using maximum likelihood, and then calculate their corresponding MDL values using (\ref{eqn:mdladditive}) or~(\ref{eqn:mdladditiverobust}).  The model that gives the smallest MDL value is taken as the best fitting model.

\section{Empirical Properties}
\label{sec:empirical}
This section investigates the empirical properties of the proposed work via numerical experiments and a real data example.

\subsection{Simulation: Linear Regression}
\label{sec:simlinear}
Following the settings in \cite{fan2012variance}, the data were generated with the model
$$y_i = b(x_{i1}+\ldots+x_{id})+\varepsilon_i$$ for $i = 1, \ldots, n$,
where the coefficient $b$ controls the signal-to-noise ratio.  The $\bm{x}_i$'s are standard normal variables with the correlation between $\bm{x}_i$ and $\bm{x}_j$ set to be $\rho^{|i-j|}$ with $\rho=0.5$.  The number of observations was $n$, and the number of predictors was $p$, where only the first $d$ are significant.  Five combinations of $(n, p, d)$ were used: $(100, 1000, 3)$, $(200, 3000, 5)$, $(300, 10000, 8)$, $(200, 100000, 5)$ and $(300, 200000, 8)$.  For each of these 5 combinations, 3 values of $b$ were used: $b = 2/\sqrt{d}, 3/\sqrt{d}$ and $5/\sqrt{d}$.  For the error term $\bm{\varepsilon}$, 4 distributions were used: $N(0,1)$, $\text{Laplace}(0,1)$, $t_3$ and a Gaussian mixture with two components: $95\% N(0,1)$ and $5\% N(0,7^2)$.  The last one represents the situation where roughly $5\%$ of the observations are outliers.  Therefore, a total of $5\times3\times4=60$ experimental configurations were considered.  The number of repetitions for each experimental configuration was 500. 

For each generated data set, six methods were applied to select a best fitting model:
\begin{enumerate}
	\item MDL: the MDL method proposed in Section~\ref{sec:linear},
	\item RobustMDL: the robust version proposed in Section~\ref{sec:robust},
	\item RLARS: the robust LARS method of \cite{khan2007robust},
	\item LAD-LASSO: the least absolute deviation lasso of \cite{wang2007robust}
	\item SparseLTS: the sparse least trimmed squares method of \cite{alfons2013sparse}, and
	\item WELSH: the adaptive welsh estimators of \cite{amato2021penalised}. Since this method is quite computationally expensive, we only applied it in the cases with $n=100$ and $p=1000$.
\end{enumerate}
To evaluate the performances of different methods on variable selection, we calculated the false negative error of selection (FN) and the false positive error of selection (FP), defined respectively as
$$
\text{FN} = \#\text{ of }\{i:\beta_i\ne0\ \&\ \hat{\beta}_i=0\}
$$
and
\begin{equation*}
	\text{FP} = \#\text{ of }\{i:\beta_i=0\ \&\ \hat{\beta}_i\ne0\}.
\end{equation*}
Note that FN can measure the ability of detecting true significant variables while FP measure the ability of removing those insignificant ones. When FN and FP are both 0, the method can detect all true significant variables while exclude those insignificant ones.
We also calculated the F1 score and mean squared error (MSE) between the estimated and true signal $\bm{X\beta}$.

The FN, FP, F1 score and MSE values for the 60 different experimental configurations obtained by the five methods are summarized in Tables~\ref{tab:1} to~\ref{tab:last}.  When considering computational speeds and performances, it seems that RobustMDL is the preferred method.

\subsection{Simulation: Nonparametric Additive Models}
For nonparametric additive models defined as~(\ref{model:additive}), we set $n=400$ and $p=1000$.  Only the first four $f_j$'s are significant:
\begin{eqnarray*}
	f_1(x) &= &5x,\\
	f_2(x) & = & 3(2x-1)^2,\\
	f_3(x) &=& 4\text{sin}(2\pi x)/\{2-\text{sin}(2\pi x)\},\\
	f_4(x) & = & 6\{0.1\text{sin}(2\pi x)+0.2\text{cos}(2\pi x)+0.3\text{sin}^2(2\pi x)\\
	& &+0.4\text{cos}^3(2\pi x)+0.5\text{sin}^3(2\pi x)\},\\
	f_j(x) &= & 0 \quad \text{for} \quad 5\le j\le p.
\end{eqnarray*}
The errors $\varepsilon_i$ were generated from four distributions: $N(0,1)$, $\text{Laplace}(0,1)$, $t_5$ and a Gaussian mixture with $95\% N(0,1)$ and $5\% N(0,5^2)$.  For each $i$, the $x_{ij}$'s were generated from
\[
x_{ij} =\begin{cases} (\omega_{ij}+tu_i)/(1+t) & \text{ for }j=1,\ldots,4 \\
	(\omega_{ij}+tk_i)/(1+t) & \text{ for }j=5,\ldots,p,
\end{cases}
\]
where $\omega_{i1},\ldots,\omega_{ip},u_i,k_i$ were i.i.d~Uniform(0,1).  The parameter $t$ controls the correlation among the predictors, and we used $t=(0,1)$ in our simulation.  Also, we used the cubic B-spline with six evenly spaced knots for all the function $f_j$'s; that is, we used $d_n = 9$ basis functions to approximate each $f_j$.  Therefore, in total there were 8 experimental configurations, and the number of replications in each configuration was 500.  For each replication, we obtained MDL and RobustMDL estimates by minimizing~(\ref{eqn:mdladditive}) and~(\ref{eqn:mdladditiverobust}).  The results are summarized in Table~\ref{tab:hd}.  We only tested these two methods as we are not aware of any other method that performs robust fitting for high-dimensional additive models.

From the simulation results, Robust\text{MDL} gave better performances in terms of FN and MSE, and provided similar results in FP as with MDL.

\subsection{Real Data Example}

This subsection presents a real data analysis on a riboflavin (vitamin $B_2$) production data set which is available in Supplementary Section A.1 of \cite{buhlmann2013statistical}.  The response variable is the logarithm of the riboflavin production rate in Bacillus subtilis for $n=71$ samples while there are $p=4,088$ covariates measuring the logarithm of the expression level of the $p=4,088$ genes.  Linear models were used in \cite{buhlmann2013statistical} and \cite{javanmard2014confidence} to detect significant genes that potentially affect riboflavin production.  The gene {\fontfamily{cmss}\selectfont YXLD-at} was located by \cite{buhlmann2013statistical} while the two genes {\fontfamily{cmss}\selectfont YXLD-at} and {\fontfamily{cmss}\selectfont YXLE-at} were identified by \cite{javanmard2014confidence} as significant.  Here in addition to fitting a linear model, we also fit a nonparametric additive model to select the significant genes.

Following \cite{buhlmann2013statistical}, we first adopted a screening procedure and only used the $100$ genes with the largest empirical variances.  We then applied the proposed Robust\text{MDL} method to fit a linear model to the screened data set.  The resulting model identified five genes as significant: {\fontfamily{cmss}\selectfont YCKE-at}, {\fontfamily{cmss}\selectfont YXLD-at}, {\fontfamily{cmss}\selectfont YDAR-at}, {\fontfamily{cmss}\selectfont XHLA-at} and {\fontfamily{cmss}\selectfont YOAB-at}.

We also fitted a nonparametric additive model using RobustMDL to the screened data set, which suggested that {\fontfamily{cmss}\selectfont YXLD-at} and {\fontfamily{cmss}\selectfont PHRI-r-at} are significant.  In other words both the fitted RobustMDL linear model and nonparametric additive model were capable of detecting {\fontfamily{cmss}\selectfont YXLD-at}, which was considered significant in most previous analyses on this data set.

\section{Concluding Remarks}
\label{sec:conclude}
The MDL principle has long been adopted by researchers in different fields to perform various estimation tasks.  In this paper we extended its use to some ``large $p$ small $n$'' problems, including high-dimensional linear regression, nonparametric additive models, as well as their robust counterparts.  As can be seen from above, one attractiveness of the MDL principle is that it can be applied to handle such problems in a natural manner; that is, by incorporating the code length of the additional parameters that are needed to specify the models.

The work presented above focused on the so-called two-part code version of MDL.  It will be interesting to develop similar methods using the newer versions of MDL, such as normalized maximum likelihood \cite{grunwald2007, rissanen2007information}, which might lead to further improved performances.



\begin{table}[ht!]
    {\small
        \begin{center}
            \caption{The FN, FP and MSE values for the methods compared in Section~\protect\ref{sec:simlinear} for those experimental settings with ($n,p$) = (100,1000) and $\varepsilon_i\sim N(0,1)$.}
            \begin{tabular}{c c c c c c c}
                \hline
                ($d,b$) & method & FN & FP & F1 & MSE & time(s) \\
                \hline
                \multirow{6}{3em}{(3,$\frac{2}{\sqrt{3}}$)} & \text{MDL} & 0.00 & 0.01 & 1.00 & 0.04 & 0.05\\
                & Robust\text{MDL} & 0.00 & 0.09 & 0.99 & 0.05 & 0.04\\
                & RLARS & 0.00 & 3.84 & 0.65 & 0.25 & 0.14\\
                & LAD-lasso & 0.00 & 0.79 & 0.90 & 0.35 & 1.06\\
                & SparseLTS & 0.00 & 8.71 & 0.55 & 0.40 & 2.93\\
                & Welsh & 0.90 & 0.19 & 0.68 & 2.15 & 780.46\\
                \hline
                \multirow{6}{3em}{(3,$\frac{3}{\sqrt{3}}$)} & \text{MDL} & 0.00 & 0.02 & 1.00 & 0.04 & 0.05\\
                & Robust\text{MDL} & 0.00 & 0.09 & 0.99 & 0.05 & 0.04\\
                & RLARS & 0.00 & 3.28 & 0.69 & 0.20 & 0.15\\
                & LAD-lasso & 0.00 & 0.78 & 0.90 & 0.31 & 1.14\\
                & SparseLTS & 0.00 & 3.59 & 0.73 & 0.37 & 2.60\\
                & Welsh & 0.58 & 0.15 & 0.79 & 3.22 & 792.34\\
                \hline
                \multirow{6}{3em}{(3,$\frac{5}{\sqrt{3}}$)} & \text{MDL} & 0.00 & 0.01 & 1.00 & 0.04 & 0.05\\
                & Robust\text{MDL} & 0.00 & 0.04 & 0.99 & 0.04 & 0.04\\
                & RLARS & 0.00 & 2.30 & 0.77 & 0.14 & 0.15\\
                & LAD-lasso & 0.00 & 0.70 & 0.91 & 0.31 & 1.26\\
                & SparseLTS & 0.00 & 2.23 & 0.76 & 0.34 & 2.37\\
                & Welsh & 0.51 & 0.34 & 0.80 & 7.73 & 786.17\\
                \hline
            \end{tabular}
            \label{tab:1}
        \end{center}
    }
\end{table}

\begin{table}[ht!]
    {\small
        \begin{center}
            \caption{Similar to Table~\protect\ref{tab:1} but for settings with ($n,p$) = $(100,1000)$ and $\varepsilon_i\sim\text{Laplace}(0,1)$.}
            \begin{tabular}{c c c c c c c}
                \hline
                ($d,b$) & method & FN & FP & F1 & MSE & time(s) \\
                \hline
                \multirow{6}{3em}{(3,$\frac{2}{\sqrt{3}}$)} & \text{MDL} & 0.03 & 0.05 & 0.99 & 0.12 & 0.05\\
                & Robust\text{MDL} & 0.00 & 0.05 & 0.99 & 0.09 & 0.04\\
                & RLARS & 0.00 & 1.99 & 0.81 & 0.24 & 0.14\\
                & LAD-lasso & 0.00 & 0.47 & 0.94 & 0.45 & 1.10\\
                & SparseLTS & 0.00 & 15.24 & 0.35 & 0.45 & 3.28\\
                & Welsh & 0.63 & 0.13 & 0.78 & 1.56 & 1045.80\\
                \hline
                \multirow{6}{3em}{(3,$\frac{3}{\sqrt{3}}$)} & \text{MDL} & 0.00 & 0.02 & 1.00 & 0.09 & 0.05\\
                & Robust\text{MDL} & 0.00 & 0.02 & 1.00 & 0.09 & 0.04\\
                & RLARS & 0.00 & 1.64 & 0.84 & 0.20 & 0.14\\
                & LAD-lasso & 0.00 & 0.47 & 0.94 & 0.45 & 1.15\\
                & SparseLTS & 0.00 & 7.79 & 0.53 & 0.45 & 2.74\\
                & Welsh & 0.39 & 0.07 & 0.87 & 2.12 & 1034.46\\
                \hline
                \multirow{6}{3em}{(3,$\frac{5}{\sqrt{3}}$)} & \text{MDL} & 0.00 & 0.01 & 1.00 & 0.09 & 0.05\\
                & Robust\text{MDL} & 0.00 & 0.01 & 1.00 & 0.09 & 0.04\\
                & RLARS & 0.00 & 1.22 & 0.87 & 0.16 & 0.14\\
                & LAD-lasso & 0.00 & 0.48 & 0.94 & 0.43 & 1.26\\
                & SparseLTS & 0.00 & 4.11 & 0.64 & 0.49 & 2.40\\
                & Welsh & 0.37 & 0.18 & 0.86 & 5.26 & 1031.58\\
                \hline
            \end{tabular}
        \end{center}
    }
\end{table}    

\begin{table}[ht!]
    {\small
        \begin{center}
            \caption{Similar to Table~\protect\ref{tab:1} but for settings with ($n,p$) = $(100,1000)$ and $\varepsilon_i\sim t_3$.}
            \begin{tabular}{c c c c c c c}
                \hline
                ($d,b$) & method & FN & FP & F1 & MSE & time(s) \\
                \hline
                \multirow{6}{3em}{(3,$\frac{2}{\sqrt{3}}$)} & \text{MDL} & 0.23 & 0.06 & 0.94 & 0.35 & 0.05\\
                & Robust\text{MDL} & 0.02 & 0.05 & 0.99 & 0.17 & 0.04\\
                & RLARS & 0.01 & 2.45 & 0.76 & 0.32 & 0.15\\
                & LAD-lasso & 0.04 & 0.36 & 0.94 & 0.63 & 1.18\\
                & SparseLTS & 0.01 & 15.70 & 0.35 & 0.55 & 3.56\\
                & Welsh & 0.70 & 0.24 & 0.75 & 1.70 & 1206.73\\
                \hline
                \multirow{6}{3em}{(3,$\frac{3}{\sqrt{3}}$)} & \text{MDL} & 0.00 & 0.03 & 1.00 & 0.11 & 0.04\\
                & Robust\text{MDL} & 0.00 & 0.00 & 1.00 & 0.09 & 0.04\\
                & RLARS & 0.00 & 1.81 & 0.82 & 0.22 & 0.13\\
                & LAD-lasso & 0.00 & 0.38 & 0.95 & 0.49 & 1.11\\
                & SparseLTS & 0.00 & 8.03 & 0.52 & 0.41 & 2.62\\
                & Welsh & 0.33 & 0.14 & 0.88 & 1.80 & 1013.85\\
                \hline
                \multirow{6}{3em}{(3,$\frac{5}{\sqrt{3}}$)} & \text{MDL} & 0.00 & 0.03 & 0.99 & 0.17 & 0.05\\
                & Robust\text{MDL} & 0.00 & 0.01 & 1.00 & 0.13 & 0.04\\
                & RLARS & 0.00 & 1.43 & 0.85 & 0.18 & 0.14\\
                & LAD-lasso & 0.00 & 0.40 & 0.95 & 0.53 & 1.27\\
                & SparseLTS & 0.00 & 4.39 & 0.62 & 0.48 & 2.46\\
                & Welsh & 0.39 & 0.18 & 0.86 & 5.61 & 1123.60\\
                \hline
            \end{tabular}
        \end{center}
    }
\end{table}    

\begin{table}[ht!]
    {\small
        \begin{center}
            \caption{Similar to Table~\protect\ref{tab:1} but for settings with ($n,p$) = $(100,1000)$ and with outliers $95\% N(0,1)\ \&\ 5\% N(0,7^2)$.}
            \begin{tabular}{c c c c c c c}
                \hline
                ($d,b$) & method & FN & FP & F1 & MSE & time(s) \\
                \hline
                \multirow{6}{3em}{(3,$\frac{2}{\sqrt{3}}$)} & \text{MDL} & 0.39 & 0.03 & 0.90 & 0.49 & 0.05\\
                & Robust\text{MDL} & 0.00 & 0.03 & 1.00 & 0.15 & 0.04\\
                & RLARS & 0.00 & 2.87 & 0.73 & 0.21 & 0.15\\
                & LAD-lasso & 0.01 & 0.26 & 0.96 & 0.49 & 1.11\\
                & SparseLTS & 0.00 & 9.83 & 0.51 & 0.39 & 3.37\\
                & Welsh & 0.49 & 0.18 & 0.82 & 1.23 & 1165.00\\
                \hline
                \multirow{6}{3em}{(3,$\frac{3}{\sqrt{3}}$)} & \text{MDL} & 0.01 & 0.03 & 0.99 & 0.17 & 0.04\\
                & Robust\text{MDL} & 0.00 & 0.02 & 1.00 & 0.13 & 0.04\\
                & RLARS & 0.00 & 2.13 & 0.79 & 0.16 & 0.14\\
                & LAD-lasso & 0.00 & 0.24 & 0.97 & 0.47 & 1.08\\
                & SparseLTS & 0.00 & 4.89 & 0.65 & 0.35 & 2.63\\
                & Welsh & 0.45 & 0.14 & 0.84 & 2.42 & 1069.91\\
                \hline
                \multirow{6}{3em}{(3,$\frac{5}{\sqrt{3}}$)} & \text{MDL} & 0.00 & 0.01 & 1.00 & 0.14 & 0.05\\
                & Robust\text{MDL} & 0.00 & 0.01 & 1.00 & 0.14 & 0.04\\
                & RLARS & 0.00 & 1.74 & 0.83 & 0.13 & 0.14\\
                & LAD-lasso & 0.00 & 0.29 & 0.96 & 0.45 & 1.20\\
                & SparseLTS & 0.00 & 2.87 & 0.71 & 0.33 & 2.41\\
                & Welsh & 0.42 & 0.29 & 0.83 & 6.29 & 1139.08\\
                \hline
            \end{tabular}
        \end{center}
    }
\end{table}

\begin{table}[ht!]
    {\small
        \begin{center}
            \caption{Similar to Table~\protect\ref{tab:1} but for settings with ($n,p$) = $(200,3000)$ and $\varepsilon_i\sim N(0,1)$.}
            \begin{tabular}{c c c c c c c}
                \hline
                ($d,b$) & method & FN & FP & F1 & MSE & time(s) \\
                \hline
                \multirow{6}{3em}{(5,$\frac{2}{\sqrt{5}}$)} & \text{MDL} & 0.00 & 0.03 & 1.00 & 0.03 & 0.11\\
                & Robust\text{MDL} & 0.00 & 0.12 & 0.99 & 0.04 & 0.10\\
                & RLARS & 0.00 & 2.87 & 0.79 & 0.14 & 0.28\\
                & LAD-lasso & 0.00 & 0.81 & 0.93 & 0.27 & 4.79\\
                & SparseLTS & 0.00 & 3.66 & 0.79 & 0.28 & 18.86\\
                \hline
                \multirow{6}{3em}{(5,$\frac{3}{\sqrt{5}}$)} & \text{MDL} & 0.00 & 0.01 & 1.00 & 0.03 & 0.11\\
                & Robust\text{MDL} & 0.00 & 0.06 & 0.99 & 0.03 & 0.10\\
                & RLARS & 0.00 & 2.37 & 0.82 & 0.11 & 0.28\\
                & LAD-lasso & 0.00 & 0.88 & 0.93 & 0.26 & 5.03\\
                & SparseLTS & 0.00 & 5.53 & 0.70 & 0.27 & 17.79\\
                \hline
                \multirow{6}{3em}{(5,$\frac{5}{\sqrt{5}}$)} & \text{MDL} & 0.00 & 0.00 & 1.00 & 0.03 & 0.11\\
                & Robust\text{MDL} & 0.00 & 0.02 & 1.00 & 0.03 & 0.11\\
                & RLARS & 0.00 & 1.64 & 0.87 & 0.07 & 0.28\\
                & LAD-lasso & 0.00 & 0.88 & 0.93 & 0.25 & 5.89\\
                & SparseLTS & 0.00 & 0.59 & 0.95 & 0.35 & 17.57\\
                \hline
            \end{tabular}
        \end{center}
    }
\end{table}    

\begin{table}[ht!]
    {\small
        \begin{center}
            \caption{Similar to Table~\protect\ref{tab:1} but for settings with ($n,p$) = $(200,3000)$ and $\varepsilon_i\sim\text{Laplace}(0,1)$.}
            \begin{tabular}{c c c c c c c}
                \hline
                ($d,b$) & method & FN & FP & F1 & MSE & time(s) \\
                \hline
                \multirow{6}{3em}{(5,$\frac{2}{\sqrt{5}}$)} & \text{MDL} & 0.02 & 0.03 & 0.99 & 0.07 & 0.11\\
                & Robust\text{MDL} & 0.00 & 0.03 & 1.00 & 0.06 & 0.10\\
                & RLARS & 0.00 & 1.20 & 0.91 & 0.12 & 0.27\\
                & LAD-lasso & 0.00 & 0.50 & 0.96 & 0.35 & 5.08\\
                & SparseLTS & 0.00 & 11.40 & 0.59 & 0.33 & 19.91\\
                \hline
                \multirow{6}{3em}{(5,$\frac{3}{\sqrt{5}}$)} & \text{MDL} & 0.00 & 0.01 & 1.00 & 0.06 & 0.11\\
                & Robust\text{MDL} & 0.00 & 0.01 & 1.00 & 0.06 & 0.10\\
                & RLARS & 0.00 & 0.91 & 0.93 & 0.10 & 0.27\\
                & LAD-lasso & 0.00 & 0.50 & 0.96 & 0.34 & 5.30\\
                & SparseLTS & 0.00 & 8.15 & 0.63 & 0.34 & 18.44\\
                \hline
                \multirow{6}{3em}{(5,$\frac{5}{\sqrt{5}}$)} & \text{MDL} & 0.00 & 0.00 & 1.00 & 0.06 & 0.11\\
                & Robust\text{MDL} & 0.00 & 0.00 & 1.00 & 0.06 & 0.10\\
                & RLARS & 0.00 & 0.64 & 0.95 & 0.08 & 0.27\\
                & LAD-lasso & 0.00 & 0.51 & 0.96 & 0.31 & 6.14\\
                & SparseLTS & 0.00 & 2.28 & 0.84 & 0.41 & 17.62\\
                \hline
            \end{tabular}
        \end{center}
    }
\end{table}    

\begin{table}[ht!]
    {\small
        \begin{center}
            \caption{Similar to Table~\protect\ref{tab:1} but for settings with ($n,p$) = $(200,3000)$ and $\varepsilon_i\sim t_3$.}
            \begin{tabular}{c c c c c c c}
                \hline
                ($d,b$) & method & FN & FP & F1 & MSE & time(s) \\
                \hline
                \multirow{6}{3em}{(5,$\frac{2}{\sqrt{5}}$)} & \text{MDL} & 0.17 & 0.05 & 0.97 & 0.19 & 0.11\\
                & Robust\text{MDL} & 0.00 & 0.05 & 1.00 & 0.10 & 0.10\\
                & RLARS & 0.00 & 1.67 & 0.88 & 0.15 & 0.27\\
                & LAD-lasso & 0.00 & 0.42 & 0.96 & 0.44 & 5.30\\
                & SparseLTS & 0.00 & 10.68 & 0.60 & 0.36 & 20.06\\
                \hline
                \multirow{6}{3em}{(5,$\frac{3}{\sqrt{5}}$)} & \text{MDL} & 0.00 & 0.05 & 1.00 & 0.11 & 0.11\\
                & Robust\text{MDL} & 0.00 & 0.03 & 1.00 & 0.09 & 0.11\\
                & RLARS & 0.00 & 1.15 & 0.91 & 0.11 & 0.28\\
                & LAD-lasso & 0.00 & 0.44 & 0.96 & 0.42 & 5.68\\
                & SparseLTS & 0.00 & 8.21 & 0.63 & 0.36 & 18.82\\
                \hline
                \multirow{6}{3em}{(5,$\frac{5}{\sqrt{5}}$)} & \text{MDL} & 0.00 & 0.03 & 1.00 & 0.10 & 0.11\\
                & Robust\text{MDL} & 0.00 & 0.02 & 1.00 & 0.09 & 0.10\\
                & RLARS & 0.00 & 0.77 & 0.94 & 0.08 & 0.27\\
                & LAD-lasso & 0.00 & 0.42 & 0.96 & 0.40 & 6.27\\
                & SparseLTS & 0.00 & 2.54 & 0.82 & 0.41 & 17.83\\
                \hline
            \end{tabular}
        \end{center}
    }
\end{table}    

\begin{table}[ht!]
    {\small
        \begin{center}
            \caption{Similar to Table~\protect\ref{tab:1} but for settings with ($n,p$) = $(200,3000)$ and with outliers $95\% N(0,1)\ \&\ 5\% N(0,7^2)$.}
            \begin{tabular}{c c c c c c c}
                \hline
                ($d,b$) & method & FN & FP & F1 & MSE & time(s) \\
                \hline
                \multirow{6}{3em}{(5,$\frac{2}{\sqrt{5}}$)} & \text{MDL} & 0.30 & 0.03 & 0.96 & 0.27 & 0.11\\
                & Robust\text{MDL} & 0.00 & 0.02 & 1.00 & 0.11 & 0.11\\
                & RLARS & 0.00 & 1.97 & 0.85 & 0.10 & 0.29\\
                & LAD-lasso & 0.00 & 0.22 & 0.98 & 0.37 & 5.17\\
                & SparseLTS & 0.00 & 4.73 & 0.75 & 0.27 & 20.22\\
                \hline
                \multirow{6}{3em}{(5,$\frac{3}{\sqrt{5}}$)} & \text{MDL} & 0.00 & 0.02 & 1.00 & 0.12 & 0.11\\
                & Robust\text{MDL} & 0.00 & 0.01 & 1.00 & 0.11 & 0.10\\
                & RLARS & 0.00 & 1.48 & 0.89 & 0.08 & 0.27\\
                & LAD-lasso & 0.00 & 0.21 & 0.98 & 0.36 & 5.12\\
                & SparseLTS & 0.00 & 6.41 & 0.67 & 0.26 & 18.36\\
                \hline
                \multirow{6}{3em}{(5,$\frac{5}{\sqrt{5}}$)} & \text{MDL} & 0.00 & 0.01 & 1.00 & 0.11 & 0.11\\
                & Robust\text{MDL} & 0.00 & 0.00 & 1.00 & 0.11 & 0.11\\
                & RLARS & 0.00 & 1.15 & 0.91 & 0.06 & 0.28\\
                & LAD-lasso & 0.00 & 0.21 & 0.98 & 0.34 & 5.95\\
                & SparseLTS & 0.00 & 0.87 & 0.93 & 0.34 & 17.86\\
                \hline
            \end{tabular}
        \end{center}
    }
\end{table}    

\begin{table}[ht!]
    {\small
        \begin{center}
            \caption{Similar to Table~\protect\ref{tab:1} but for settings with ($n,p$) = $(300,10000)$ and $\varepsilon_i\sim N(0,1)$.}
            \begin{tabular}{c c c c c c c}
                \hline
                ($d,b$) & method & FN & FP & F1 & MSE & time(s) \\
                \hline
                \multirow{6}{3em}{(8,$\frac{2}{\sqrt{8}}$)} & \text{MDL} & 0.00 & 0.01 & 1.00 & 0.03 & 0.29\\
                & Robust\text{MDL} & 0.00 & 0.08 & 1.00 & 0.03 & 0.29\\
                & RLARS & 0.00 & 1.10 & 0.94 & 0.07 & 0.54\\
                & LAD-lasso & 0.00 & 1.34 & 0.93 & 0.30 & 15.16\\
                & SparseLTS & 0.00 & 3.74 & 0.84 & 0.31 & 64.98\\
                \hline
                \multirow{6}{3em}{(8,$\frac{3}{\sqrt{8}}$)} & \text{MDL} & 0.00 & 0.01 & 1.00 & 0.03 & 0.29\\
                & Robust\text{MDL} & 0.00 & 0.04 & 1.00 & 0.03 & 0.29\\
                & RLARS & 0.00 & 0.84 & 0.95 & 0.05 & 0.53\\
                & LAD-lasso & 0.00 & 1.34 & 0.93 & 0.29 & 15.84\\
                & SparseLTS & 0.00 & 8.73 & 0.68 & 0.27 & 62.71\\
                \hline
                \multirow{6}{3em}{(8,$\frac{5}{\sqrt{8}}$)} & \text{MDL} & 0.00 & 0.00 & 1.00 & 0.03 & 0.32\\
                & Robust\text{MDL} & 0.00 & 0.02 & 1.00 & 0.03 & 0.32\\
                & RLARS & 0.00 & 0.61 & 0.97 & 0.04 & 0.58\\
                & LAD-lasso & 0.00 & 1.28 & 0.93 & 0.27 & 19.25\\
                & SparseLTS & 0.00 & 0.37 & 0.98 & 0.41 & 63.62\\
                \hline
            \end{tabular}
        \end{center}
    }
\end{table}    

\begin{table}[ht!]
    {\small
        \begin{center}
            \caption{Similar to Table~\protect\ref{tab:1} but for settings with ($n,p$) = $(300,10000)$ and $\varepsilon_i\sim\text{Laplace}(0,1)$.}
            \begin{tabular}{c c c c c c c}
                \hline
                ($d,b$) & method & FN & FP & F1 & MSE & time(s) \\
                \hline
                \multirow{6}{3em}{(8,$\frac{2}{\sqrt{8}}$)} & \text{MDL} & 0.04 & 0.04 & 0.99 & 0.08 & 0.29\\
                & Robust\text{MDL} & 0.00 & 0.01 & 1.00 & 0.06 & 0.29\\
                & RLARS & 0.00 & 0.57 & 0.97 & 0.07 & 0.52\\
                & LAD-lasso & 0.00 & 0.82 & 0.95 & 0.40 & 16.27\\
                & SparseLTS & 0.00 & 14.59 & 0.62 & 0.36 & 67.98\\
                \hline
                \multirow{6}{3em}{(8,$\frac{3}{\sqrt{8}}$)} & \text{MDL} & 0.00 & 0.02 & 1.00 & 0.06 & 0.30\\
                & Robust\text{MDL} & 0.00 & 0.00 & 1.00 & 0.06 & 0.30\\
                & RLARS & 0.00 & 0.44 & 0.97 & 0.06 & 0.54\\
                & LAD-lasso & 0.00 & 0.88 & 0.95 & 0.38 & 17.45\\
                & SparseLTS & 0.00 & 12.83 & 0.61 & 0.35 & 64.56\\
                \hline
                \multirow{6}{3em}{(8,$\frac{5}{\sqrt{8}}$)} & \text{MDL} & 0.00 & 0.00 & 1.00 & 0.06 & 0.30\\
                & Robust\text{MDL} & 0.00 & 0.00 & 1.00 & 0.06 & 0.30\\
                & RLARS & 0.00 & 0.34 & 0.98 & 0.06 & 0.53\\
                & LAD-lasso & 0.00 & 0.83 & 0.95 & 0.35 & 19.64\\
                & SparseLTS & 0.00 & 2.36 & 0.88 & 0.49 & 62.71\\
                \hline
            \end{tabular}
        \end{center}
    }
\end{table}    

\begin{table}[ht!]
    {\small
        \begin{center}
            \caption{Similar to Table~\protect\ref{tab:1} but for settings with ($n,p$) = $(300,10000)$ and $\varepsilon_i\sim t_3$.}
            \begin{tabular}{c c c c c c c}
                \hline
                ($d,b$) & method & FN & FP & F1 & MSE & time(s) \\
                \hline
                \multirow{6}{3em}{(8,$\frac{2}{\sqrt{8}}$)} & \text{MDL} & 0.42 & 0.02 & 0.97 & 0.23 & 0.30\\
                & Robust\text{MDL} & 0.02 & 0.02 & 1.00 & 0.10 & 0.30\\
                & RLARS & 0.01 & 0.71 & 0.96 & 0.09 & 0.54\\
                & LAD-lasso & 0.01 & 0.61 & 0.96 & 0.50 & 16.98\\
                & SparseLTS & 0.01 & 12.33 & 0.65 & 0.38 & 69.30\\
                \hline
                \multirow{6}{3em}{(8,$\frac{3}{\sqrt{8}}$)} & \text{MDL} & 0.04 & 0.02 & 1.00 & 0.14 & 0.31\\
                & Robust\text{MDL} & 0.01 & 0.01 & 1.00 & 0.10 & 0.31\\
                & RLARS & 0.01 & 0.58 & 0.97 & 0.08 & 0.54\\
                & LAD-lasso & 0.01 & 0.68 & 0.96 & 0.50 & 18.01\\
                & SparseLTS & 0.01 & 11.77 & 0.64 & 0.39 & 65.30\\
                \hline
                \multirow{6}{3em}{(8,$\frac{5}{\sqrt{8}}$)} & \text{MDL} & 0.01 & 0.01 & 1.00 & 0.16 & 0.30\\
                & Robust\text{MDL} & 0.00 & 0.00 & 1.00 & 0.11 & 0.30\\
                & RLARS & 0.00 & 0.44 & 0.98 & 0.06 & 0.54\\
                & LAD-lasso & 0.02 & 0.68 & 0.96 & 0.56 & 20.15\\
                & SparseLTS & 0.00 & 2.58 & 0.87 & 0.47 & 62.98\\
                \hline
            \end{tabular}
        \end{center}
    }
\end{table}

\begin{table}[ht!]
    {\small
        \begin{center}
            \caption{Similar to Table~\protect\ref{tab:1} but for settings with ($n,p$) = $(300,10000)$ and with outliers $95\% N(0,1)\ \&\ 5\% N(0,7^2)$.}
            \begin{tabular}{c c c c c c c}
                \hline
                ($d,b$) & method & FN & FP & F1 & MSE & time(s) \\
                \hline
                \multirow{6}{3em}{(8,$\frac{2}{\sqrt{8}}$)} & \text{MDL} & 0.71 & 0.01 & 0.95 & 0.33 & 0.30\\
                & Robust\text{MDL} & 0.01 & 0.01 & 1.00 & 0.11 & 0.30\\
                & RLARS & 0.00 & 0.80 & 0.95 & 0.06 & 0.54\\
                & LAD-lasso & 0.00 & 0.38 & 0.98 & 0.43 & 15.71\\
                & SparseLTS & 0.00 & 5.43 & 0.79 & 0.29 & 68.05\\
                \hline
                \multirow{6}{3em}{(8,$\frac{3}{\sqrt{8}}$)} & \text{MDL} & 0.01 & 0.01 & 1.00 & 0.11 & 0.31\\
                & Robust\text{MDL} & 0.00 & 0.00 & 1.00 & 0.11 & 0.31\\
                & RLARS & 0.00 & 0.64 & 0.96 & 0.05 & 0.55\\
                & LAD-lasso & 0.00 & 0.36 & 0.98 & 0.41 & 17.31\\
                & SparseLTS & 0.00 & 10.38 & 0.64 & 0.25 & 65.13\\
                \hline
                \multirow{6}{3em}{(8,$\frac{5}{\sqrt{8}}$)} & \text{MDL} & 0.00 & 0.00 & 1.00 & 0.11 & 0.30\\
                & Robust\text{MDL} & 0.00 & 0.00 & 1.00 & 0.11 & 0.30\\
                & RLARS & 0.00 & 0.42 & 0.98 & 0.04 & 0.54\\
                & LAD-lasso & 0.00 & 0.35 & 0.98 & 0.40 & 19.18\\
                & SparseLTS & 0.00 & 0.77 & 0.96 & 0.40 & 62.94\\
                \hline
            \end{tabular}
            \label{tab:16}
        \end{center}
    }
\end{table}    

\begin{table}[ht!]
    {\small
        \begin{center}
            \caption{Similar to Table~\protect\ref{tab:1} but for settings with ($n,p$) = $(200,10000)$ and $\varepsilon_i\sim N(0,1)$.}
            \begin{tabular}{c c c c c c c}
                \hline
                ($d,b$) & method & FN & FP & F1 & MSE & time(s) \\
                \hline
                \multirow{6}{3em}{(5,$\frac{2}{\sqrt{5}}$)} & \text{MDL} & 0.00 & 0.01 & 1.00 & 0.03 & 0.24\\
                & Robust\text{MDL} & 0.00 & 0.07 & 0.99 & 0.04 & 0.24\\
                & RLARS & 0.00 & 3.11 & 0.78 & 0.16 & 0.41\\
                & LAD-lasso & 0.00 & 1.05 & 0.92 & 0.31 & 5.83\\
                & SparseLTS & 0.00 & 5.13 & 0.74 & 0.31 & 19.10\\
                \hline
                \multirow{6}{3em}{(5,$\frac{3}{\sqrt{5}}$)} & \text{MDL} & 0.00 & 0.00 & 1.00 & 0.03 & 0.24\\
                & Robust\text{MDL} & 0.00 & 0.04 & 1.00 & 0.03 & 0.24\\
                & RLARS & 0.00 & 2.51 & 0.82 & 0.12 & 0.41\\
                & LAD-lasso & 0.00 & 1.01 & 0.92 & 0.30 & 5.85\\
                & SparseLTS & 0.00 & 5.77 & 0.71 & 0.31 & 18.00\\
                \hline
                \multirow{6}{3em}{(5,$\frac{5}{\sqrt{5}}$)} & \text{MDL} & 0.00 & 0.00 & 1.00 & 0.03 & 0.24\\
                & Robust\text{MDL} & 0.00 & 0.01 & 1.00 & 0.03 & 0.23\\
                & RLARS & 0.00 & 1.74 & 0.87 & 0.08 & 0.40\\
                & LAD-lasso & 0.00 & 1.07 & 0.91 & 0.28 & 6.30\\
                & SparseLTS & 0.00 & 0.88 & 0.93 & 0.35 & 17.49\\
                \hline
            \end{tabular}
        \end{center}
    }
\end{table}    

\begin{table}[ht!]
    {\small
        \begin{center}
            \caption{Similar to Table~\protect\ref{tab:1} but for settings with ($n,p$) = $(200,10000)$ and $\varepsilon_i\sim\text{Laplace}(0,1)$.}
            \begin{tabular}{c c c c c c c}
                \hline
                ($d,b$) & method & FN & FP & F1 & MSE & time(s) \\
                \hline
                \multirow{6}{3em}{(5,$\frac{2}{\sqrt{5}}$)} & \text{MDL} & 0.02 & 0.01 & 1.00 & 0.07 & 0.24\\
                & Robust\text{MDL} & 0.00 & 0.01 & 1.00 & 0.06 & 0.24\\
                & RLARS & 0.00 & 1.38 & 0.90 & 0.13 & 0.40\\
                & LAD-lasso & 0.00 & 0.60 & 0.95 & 0.42 & 6.41\\
                & SparseLTS & 0.00 & 16.30 & 0.50 & 0.36 & 20.47\\
                \hline
                \multirow{6}{3em}{(5,$\frac{3}{\sqrt{5}}$)} & \text{MDL} & 0.00 & 0.01 & 1.00 & 0.06 & 0.24\\
                & Robust\text{MDL} & 0.00 & 0.00 & 1.00 & 0.06 & 0.24\\
                & RLARS & 0.00 & 1.02 & 0.92 & 0.10 & 0.40\\
                & LAD-lasso & 0.00 & 0.60 & 0.95 & 0.39 & 6.25\\
                & SparseLTS & 0.00 & 9.74 & 0.60 & 0.39 & 18.60\\
                \hline
                \multirow{6}{3em}{(5,$\frac{5}{\sqrt{5}}$)} & \text{MDL} & 0.00 & 0.00 & 1.00 & 0.06 & 0.23\\
                & Robust\text{MDL} & 0.00 & 0.00 & 1.00 & 0.06 & 0.23\\
                & RLARS & 0.00 & 0.76 & 0.94 & 0.08 & 0.39\\
                & LAD-lasso & 0.00 & 0.60 & 0.95 & 0.36 & 6.71\\
                & SparseLTS & 0.00 & 3.36 & 0.78 & 0.43 & 17.64\\
                \hline
            \end{tabular}
        \end{center}
    }
\end{table}    

\begin{table}[ht!]
    {\small
        \begin{center}
            \caption{Similar to Table~\protect\ref{tab:1} but for settings with ($n,p$) = $(200,10000)$ and $\varepsilon_i\sim t_3$.}
            \begin{tabular}{c c c c c c c}
                \hline
                ($d,b$) & method & FN & FP & F1 & MSE & time(s) \\
                \hline
                \multirow{6}{3em}{(5,$\frac{2}{\sqrt{5}}$)} & \text{MDL} & 0.31 & 0.02 & 0.96 & 0.31 & 0.23\\
                & Robust\text{MDL} & 0.01 & 0.02 & 1.00 & 0.13 & 0.23\\
                & RLARS & 0.01 & 1.86 & 0.86 & 0.18 & 0.40\\
                & LAD-lasso & 0.02 & 0.46 & 0.96 & 0.55 & 6.61\\
                & SparseLTS & 0.01 & 17.05 & 0.50 & 0.40 & 20.86\\
                \hline
                \multirow{6}{3em}{(5,$\frac{3}{\sqrt{5}}$)} & \text{MDL} & 0.05 & 0.01 & 0.99 & 0.21 & 0.24\\
                & Robust\text{MDL} & 0.01 & 0.01 & 1.00 & 0.15 & 0.23\\
                & RLARS & 0.01 & 1.34 & 0.90 & 0.15 & 0.40\\
                & LAD-lasso & 0.01 & 0.44 & 0.96 & 0.53 & 6.45\\
                & SparseLTS & 0.01 & 9.02 & 0.63 & 0.43 & 18.90\\
                \hline
                \multirow{6}{3em}{(5,$\frac{5}{\sqrt{5}}$)} & \text{MDL} & 0.01 & 0.00 & 1.00 & 0.17 & 0.24\\
                & Robust\text{MDL} & 0.00 & 0.00 & 1.00 & 0.12 & 0.24\\
                & RLARS & 0.00 & 0.95 & 0.93 & 0.11 & 0.41\\
                & LAD-lasso & 0.01 & 0.49 & 0.96 & 0.55 & 6.91\\
                & SparseLTS & 0.00 & 3.89 & 0.75 & 0.45 & 17.88\\
                \hline
            \end{tabular}
        \end{center}
    }
\end{table}    

\begin{table}[ht!]
    {\small
        \begin{center}
            \caption{Similar to Table~\protect\ref{tab:1} but for settings with ($n,p$) = $(200,10000)$ and with outliers $95\% N(0,1)\ \&\ 5\% N(0,7^2)$.}
            \begin{tabular}{c c c c c c c}
                \hline
                ($d,b$) & method & FN & FP & F1 & MSE & time(s) \\
                \hline
                \multirow{6}{3em}{(5,$\frac{2}{\sqrt{5}}$)} & \text{MDL} & 0.45 & 0.02 & 0.94 & 0.35 & 0.24\\
                & Robust\text{MDL} & 0.00 & 0.01 & 1.00 & 0.11 & 0.23\\
                & RLARS & 0.00 & 2.07 & 0.85 & 0.11 & 0.41\\
                & LAD-lasso & 0.00 & 0.25 & 0.98 & 0.44 & 6.14\\
                & SparseLTS & 0.00 & 7.01 & 0.68 & 0.29 & 20.15\\
                \hline
                \multirow{6}{3em}{(5,$\frac{3}{\sqrt{5}}$)} & \text{MDL} & 0.01 & 0.02 & 1.00 & 0.12 & 0.23\\
                & Robust\text{MDL} & 0.00 & 0.01 & 1.00 & 0.11 & 0.23\\
                & RLARS & 0.00 & 1.63 & 0.88 & 0.09 & 0.40\\
                & LAD-lasso & 0.00 & 0.26 & 0.98 & 0.42 & 6.16\\
                & SparseLTS & 0.00 & 6.51 & 0.68 & 0.31 & 18.67\\
                \hline
                \multirow{6}{3em}{(5,$\frac{5}{\sqrt{5}}$)} & \text{MDL} & 0.00 & 0.01 & 1.00 & 0.11 & 0.24\\
                & Robust\text{MDL} & 0.00 & 0.00 & 1.00 & 0.11 & 0.24\\
                & RLARS & 0.00 & 1.23 & 0.91 & 0.07 & 0.41\\
                & LAD-lasso & 0.00 & 0.25 & 0.98 & 0.40 & 6.84\\
                & SparseLTS & 0.00 & 1.49 & 0.89 & 0.34 & 17.87\\
                \hline
            \end{tabular}
        \end{center}
    }
\end{table}    

\begin{table}[ht!]
    {\small
        \begin{center}
            \caption{Similar to Table~\protect\ref{tab:1} but for settings with ($n,p$) = $(300,20000)$ and $\varepsilon_i\sim N(0,1)$.}
            \begin{tabular}{c c c c c c c}
                \hline
                ($d,b$) & method & FN & FP & F1 & MSE & time(s) \\
                \hline
                \multirow{6}{3em}{(8,$\frac{2}{\sqrt{8}}$)} & \text{MDL} & 0.00 & 0.01 & 1.00 & 0.03 & 0.67\\
                & Robust\text{MDL} & 0.00 & 0.07 & 1.00 & 0.03 & 0.67\\
                & RLARS & 0.00 & 1.26 & 0.93 & 0.07 & 0.98\\
                & LAD-lasso & 0.00 & 1.46 & 0.92 & 0.33 & 22.71\\
                & SparseLTS & 0.00 & 4.99 & 0.80 & 0.32 & 78.90\\
                \hline
                \multirow{6}{3em}{(8,$\frac{3}{\sqrt{8}}$)} & \text{MDL} & 0.00 & 0.00 & 1.00 & 0.03 & 0.68\\
                & Robust\text{MDL} & 0.00 & 0.04 & 1.00 & 0.03 & 0.68\\
                & RLARS & 0.00 & 0.97 & 0.95 & 0.06 & 1.00\\
                & LAD-lasso & 0.00 & 1.40 & 0.93 & 0.31 & 22.94\\
                & SparseLTS & 0.00 & 8.97 & 0.68 & 0.30 & 75.60\\
                \hline
                \multirow{6}{3em}{(8,$\frac{5}{\sqrt{8}}$)} & \text{MDL} & 0.00 & 0.00 & 1.00 & 0.03 & 0.67\\
                & Robust\text{MDL} & 0.00 & 0.01 & 1.00 & 0.03 & 0.67\\
                & RLARS & 0.00 & 0.69 & 0.96 & 0.05 & 0.98\\
                & LAD-lasso & 0.00 & 1.47 & 0.92 & 0.29 & 24.46\\
                & SparseLTS & 0.00 & 0.65 & 0.96 & 0.41 & 74.36\\
                \hline
            \end{tabular}
        \end{center}
    }
\end{table}    

\begin{table}[ht!]
    {\small
        \begin{center}
            \caption{Similar to Table~\protect\ref{tab:1} but for settings with ($n,p$) = $(300,20000)$ and $\varepsilon_i\sim\text{Laplace}(0,1)$.}
            \begin{tabular}{c c c c c c c}
                \hline
                ($d,b$) & method & FN & FP & F1 & MSE & time(s) \\
                \hline
                \multirow{6}{3em}{(8,$\frac{2}{\sqrt{8}}$)} & \text{MDL} & 0.05 & 0.03 & 0.99 & 0.08 & 0.64\\
                & Robust\text{MDL} & 0.01 & 0.03 & 1.00 & 0.06 & 0.64\\
                & RLARS & 0.00 & 0.57 & 0.97 & 0.08 & 0.93\\
                & LAD-lasso & 0.00 & 0.93 & 0.95 & 0.42 & 24.54\\
                & SparseLTS & 0.00 & 18.47 & 0.57 & 0.37 & 81.99\\
                \hline
                \multirow{6}{3em}{(8,$\frac{3}{\sqrt{8}}$)} & \text{MDL} & 0.00 & 0.01 & 1.00 & 0.06 & 0.66\\
                & Robust\text{MDL} & 0.00 & 0.01 & 1.00 & 0.06 & 0.66\\
                & RLARS & 0.00 & 0.43 & 0.98 & 0.07 & 0.95\\
                & LAD-lasso & 0.00 & 0.85 & 0.95 & 0.40 & 24.52\\
                & SparseLTS & 0.00 & 13.76 & 0.60 & 0.38 & 77.14\\
                \hline
                \multirow{6}{3em}{(8,$\frac{5}{\sqrt{8}}$)} & \text{MDL} & 0.00 & 0.00 & 1.00 & 0.06 & 0.66\\
                & Robust\text{MDL} & 0.00 & 0.00 & 1.00 & 0.06 & 0.66\\
                & RLARS & 0.00 & 0.29 & 0.98 & 0.06 & 0.95\\
                & LAD-lasso & 0.00 & 0.85 & 0.95 & 0.37 & 26.72\\
                & SparseLTS & 0.00 & 3.04 & 0.85 & 0.49 & 74.17\\
                \hline
            \end{tabular}
        \end{center}
    }
\end{table}    

\begin{table}[ht!]
    {\small
        \begin{center}
            \caption{Similar to Table~\protect\ref{tab:1} but for settings with ($n,p$) = $(300,20000)$ and $\varepsilon_i\sim t_3$.}
            \begin{tabular}{c c c c c c c}
                \hline
                ($d,b$) & method & FN & FP & F1 & MSE & time(s) \\
                \hline
                \multirow{6}{3em}{(8,$\frac{2}{\sqrt{8}}$)} & \text{MDL} & 0.67 & 0.02 & 0.95 & 0.33 & 0.64\\
                & Robust\text{MDL} & 0.03 & 0.02 & 1.00 & 0.12 & 0.64\\
                & RLARS & 0.01 & 0.74 & 0.96 & 0.10 & 0.93\\
                & LAD-lasso & 0.04 & 0.67 & 0.96 & 0.57 & 25.24\\
                & SparseLTS & 0.01 & 15.96 & 0.61 & 0.41 & 83.76\\
                \hline
                \multirow{6}{3em}{(8,$\frac{3}{\sqrt{8}}$)} & \text{MDL} & 0.06 & 0.02 & 0.99 & 0.18 & 0.67\\
                & Robust\text{MDL} & 0.00 & 0.01 & 1.00 & 0.10 & 0.67\\
                & RLARS & 0.00 & 0.61 & 0.97 & 0.08 & 0.96\\
                & LAD-lasso & 0.01 & 0.69 & 0.96 & 0.54 & 25.60\\
                & SparseLTS & 0.00 & 11.58 & 0.66 & 0.42 & 77.41\\
                \hline
                \multirow{6}{3em}{(8,$\frac{5}{\sqrt{8}}$)} & \text{MDL} & 0.02 & 0.00 & 1.00 & 0.17 & 0.69\\
                & Robust\text{MDL} & 0.00 & 0.00 & 1.00 & 0.10 & 0.69\\
                & RLARS & 0.00 & 0.46 & 0.97 & 0.07 & 0.99\\
                & LAD-lasso & 0.00 & 0.66 & 0.96 & 0.48 & 27.85\\
                & SparseLTS & 0.00 & 3.65 & 0.83 & 0.47 & 75.01\\
                \hline
            \end{tabular}
        \end{center}
    }
\end{table}    

\begin{table}[ht!]
    {\small
        \begin{center}
            \caption{Similar to Table~\protect\ref{tab:1} but for settings with ($n,p$) = $(300,20000)$ and with outliers $95\% N(0,1)\ \&\ 5\% N(0,7^2)$.}
            \begin{tabular}{c c c c c c c}
                \hline
                ($d,b$) & method & FN & FP & F1 & MSE & time(s) \\
                \hline
                \multirow{6}{3em}{(8,$\frac{2}{\sqrt{8}}$)} & \text{MDL} & 0.78 & 0.01 & 0.94 & 0.36 & 0.66\\
                & Robust\text{MDL} & 0.00 & 0.01 & 1.00 & 0.11 & 0.65\\
                & RLARS & 0.00 & 0.91 & 0.95 & 0.06 & 0.95\\
                & LAD-lasso & 0.00 & 0.47 & 0.97 & 0.45 & 22.87\\
                & SparseLTS & 0.00 & 6.45 & 0.76 & 0.30 & 81.55\\
                \hline
                \multirow{6}{3em}{(8,$\frac{3}{\sqrt{8}}$)} & \text{MDL} & 0.01 & 0.01 & 1.00 & 0.11 & 0.66\\
                & Robust\text{MDL} & 0.00 & 0.00 & 1.00 & 0.10 & 0.66\\
                & RLARS & 0.00 & 0.68 & 0.96 & 0.05 & 0.96\\
                & LAD-lasso & 0.00 & 0.44 & 0.97 & 0.44 & 23.64\\
                & SparseLTS & 0.00 & 10.78 & 0.64 & 0.28 & 77.52\\
                \hline
                \multirow{6}{3em}{(8,$\frac{5}{\sqrt{8}}$)} & \text{MDL} & 0.00 & 0.01 & 1.00 & 0.11 & 0.68\\
                & Robust\text{MDL} & 0.00 & 0.00 & 1.00 & 0.10 & 0.67\\
                & RLARS & 0.00 & 0.53 & 0.97 & 0.04 & 0.98\\
                & LAD-lasso & 0.00 & 0.44 & 0.98 & 0.41 & 25.76\\
                & SparseLTS & 0.00 & 1.06 & 0.94 & 0.39 & 74.94\\
                \hline
            \end{tabular}
            \label{tab:last}
        \end{center}
    }
\end{table}

\begin{table}[ht!]
    {\small
        \begin{center}
            \caption{The SEN and SPE values for the methods compared in the nonparametric additive model experiments.}
            \begin{tabular}{c c c c c c} \hline
                error distribution & $t$ & method & FN & FP & MSE\\
                \hline
                \multirow{4}{3em}{$N(0,1)$} & \multirow{2}{0.5em}{0} & \text{MDL} & 0.01 & 0.00 & 0.93\\
                &                      & Robust\text{MDL} & 0.01 & 0.00 & 0.93\\
                & \multirow{2}{0.5em}{1} & \text{MDL} & 0.05 & 0.02 & 0.94\\
                &                      & Robust\text{MDL} & 0.05 & 0.02 & 0.94\\
                \hline
                \multirow{4}{5em}{Laplace(0,1)} & \multirow{2}{0.5em}{0} & \text{MDL} & 0.01 & 0.00 & 1.84\\
                &                      & Robust\text{MDL} & 0.01 & 0.08 & 1.83\\
                & \multirow{2}{0.5em}{1} & \text{MDL} & 0.47 & 0.02 & 1.99\\
                &                      & Robust\text{MDL} & 0.23 & 0.09 & 1.89\\
                \hline
                \multirow{4}{1em}{$t_5$} & \multirow{2}{0.5em}{0} & \text{MDL} & 0.01 & 0.00 & 1.56\\
                &                      & Robust\text{MDL} & 0.01 & 0.00 & 1.56\\
                & \multirow{2}{0.5em}{1} & \text{MDL} & 0.31 & 0.04 & 1.65\\
                &                      & Robust\text{MDL} & 0.16 & 0.08 & 1.59\\
                \hline
                \multirow{4}{3em}{Outliers} & \multirow{2}{0.5em}{0} & \text{MDL} & 0.14 & 0.00 & 3.17\\
                &                      & Robust\text{MDL} & 0.02 & 0.00 & 3.10\\
                & \multirow{2}{0.5em}{1} & \text{MDL} & 1.49 & 0.00 & 3.83\\
                &                      & Robust\text{MDL} & 0.73 & 0.17 & 3.40\\
                \hline
            \end{tabular}
            \label{tab:hd}
        \end{center}
    }
\end{table}     

\bibliographystyle{IEEEbib}
\bibliography{references}

\appendix

\section{Technical Details}
This appendix provides technical details, including the proof for Theorem~\ref{thm:1}.

\subsection{Lemmas}

{ \lemma \label{pchisq}
	
	Let $\chi_j^2$ denote a $\chi^2$ random variable with degrees of freedom $j$. If $c\rightarrow \infty$ and $\frac{J}{c}\rightarrow 0$, then
	$$P(\chi_j^2>c)=\frac{1}{\Gamma(j/2)}(c/2)^{j/2-1}\exp (-c/2)(1+o(1))$$
	uniformly for all $j\leq J$.
}

The proof can be found in \cite{Luo-Chen13} by using integration by parts.

{ \lemma \label{chisq}
	
	Let $\chi_j^2$ be a chi-square random variable with degrees of freedom $j$ and $c_j=2j[\log p+\log
	(j\log p)]$. If $p\rightarrow \infty $, then for any $J\leq p$,
	$$
	\sum_{j=1}^J\binom{p}{j}P(\chi_{j}^2>c_{j})\rightarrow 0.
	$$}

The proof is similar to the one for Lemma~\ref{pchisq} and hence is omitted.

\subsection{Proof of the Theorem~\ref{THM:1}}

\begin{proof}
	Since 
	$$ \text{MDL}(S) = \frac{n}{2}\log\left(\frac{\text{RSS}}{n}\right) + \frac{|S|}{2}\log(n) + |S|\log(p),$$
	we have
	$$\text{MDL}(S)-\text{MDL}(S_0) = T_1 + T_2,$$
	where
	$$T_1= \frac{n}{2}\log\left(\frac{\text{RSS}_S}{\text{RSS}_{S_0}}\right)$$ and
	$$T_2=(|S|-|S_0|)\log(p\sqrt{n}).$$
	
	Without loss of generality, we assume that $\sigma^2 = 1.$
	
	\noindent{\bf{Case 1}: $S_0\not\subset S$.}
	
	Denote $\mathcal{S}$ as the collection of models which hold asymptotic identifiability condition~\ref{con:1}, that is, $\mathcal{S} = \{S:|S|\le k|S_0|\}$ for some fixed $k>1$. In practice we only consider models with size comparable with the true model, so the restriction $|S|\le k|S_0|$ is imposed. Let $\mathcal{S}_j=\{S:|S|=j, S\in \mathcal{S}\}$.  Recall that $\bm{P}_S$ is the projection matrix for model $S$ and $\bm{P}_{S_0}$ is the projection matrix for the true model $S_0$. Note that
	\begin{align*}
		\text{RSS}_{S_0}&=(\bm{y}-\bm{X}_{S_0}\bm{\beta}_{S_0})^T(\bm{I}-\bm{P}_{S_0})(\bm{y}-\bm{X}_{S_0}\bm{\beta}_{S_0}) \\
		&=\bm{\varepsilon}^T(\bm{I}-\bm{P}_{S_0})\bm{\varepsilon}=\sum_{i=1}^{n-|S_0|}Z_i^2\\
		&=(n-|S_0|)(1+o_p(1))=n(1+o_p(1)),
	\end{align*}
	where $Z_i$'s are i.i.d. standard normal variables.
	
	Recall $\delta(S) = \|\bm{\mu}-\bm{P}_S\bm{\mu}\|^2$ with $\bm{\mu}=\bm{X}_{S_0}\bm{\beta}_{S_0}$. Then
	\begin{eqnarray} 
		\label{RSS_diff}
		\text{RSS}_{S}-\text{RSS}_{S_0}&=&(\bm{\mu}+\bm{\varepsilon})^T(\bm{I}-\bm{P}_S)(\bm{\mu}+\bm{\varepsilon}) \nonumber  \\
		& &-\bm{\varepsilon}^T(\bm{I}-\bm{P}_{S_0})\bm{\varepsilon} \nonumber  \\
		&=&\delta(S)+2\bm{\mu}^T(\bm{I}-\bm{P}_S)\bm{\varepsilon} \nonumber  \\
		&&-\bm{\varepsilon}^T\bm{P}_S\bm{\varepsilon}+\bm{\varepsilon}^T\bm{P}_{S_0}\bm{\varepsilon},
	\end{eqnarray}
	and $\bm{\varepsilon}^T\bm{P}_{S_0}\bm{\varepsilon}=|S_0|(1+o_p(1))$.
	
	Write the second term in (\ref{RSS_diff}) as $$\bm{\mu}^T(\bm{I}-\bm{P}_S)\bm{\varepsilon}=\sqrt{\delta(S)}Z_S,$$
	where $Z_S\sim N(0,1)$.  Then for any $S \in \mathcal{S}$,
	$$\bm{\mu}^T(\bm{I}-\bm{P}_S)\bm{\varepsilon}\leq \sqrt{\delta(S)}\max_{S\in\mathcal{S}}|Z_S|.$$
	
	Let $c_j=2j\{\log p+\log (j\log p)\}$ and $c = \max\{c_j: 1\leq j\leq k|S_0|\}$, according to Lemma \ref{chisq} we hence have
	\begin{align*}
		P(\max_{S\in\mathcal{S}}|Z_S|\geq \sqrt{c}) &= P(\max_{S\in\mathcal{S}_j,1\leq j\leq k|S_0|}|Z_S|\geq \sqrt{c}) \\
		&\leq\sum_{j=1}^{k|S_0|}\binom{p}{j}P(\chi_1^2\geq c)\\
		&\leq \sum_{j=1}^{k|S_0|}\binom{p}{j}P(\chi_j^2\geq c)\\
		&\leq \sum_{j=1}^{k|S_0|}\binom{p}{j}P(\chi_j^2\geq c_j)\rightarrow 0.
	\end{align*}
	By the identifiability condition \ref{con:1}, $\log n = o_p(\delta(S))$, we have $|\bm{\mu}^T(\bm{I}-\bm{P}_S)\bm{\varepsilon}|=\sqrt{\delta(S)O_p(k|S_0|\log p)} = o_p(\delta(S))$ uniformly over $\mathcal{S}$. 
	
	Similarly, for the third term in (\ref{RSS_diff}), as $\bm{\varepsilon}^T\bm{P}_S\bm{\varepsilon}=\chi_{|S|}^2$ we have
	\begin{align*}
		P(\max_{S\in\mathcal{S}} \bm{\varepsilon}^T\bm{P}_S\bm{\varepsilon}\geq c) & =P(\max_{S\in\mathcal{S}_j,1\leq j\leq k|S_0|} \chi_j^2 \geq c) \\
		& \leq\sum_{j=1}^{k|S_0|}\binom{p}{j}P(\chi_j^2\geq c_j)\rightarrow 0.
	\end{align*}
	Thus we have
	$$\max_{\mathcal{S}}\{\bm{\varepsilon}^T\bm{P}_S\bm{\varepsilon}\}=O_p(k|S_0|\log p)=o_p(\delta(S)).$$
	In a word, $\delta(S)$ is the dominant term in (\ref{RSS_diff}). Therefore,
	$$
	\text{RSS}_{S}-\text{RSS}_{S_0}=\delta(S)(1+o_p(1))
	$$ and
	\begin{align} \label{case1t1}
		T_1&=\frac{n}{2}\log\left(1+\frac{\text{RSS}_{S}-\text{RSS}_{S_0}}{\text{RSS}_{S_0}}\right) \nonumber \\
		&=\frac{n}{2}\log\left(1+\frac{\delta(S)(1+o_p(1))}{n}\right) \nonumber \\
		& =\frac{\delta(S)(1+o_p(1))}{2}.
	\end{align}
	
	We also have
	\begin{eqnarray}
		\label{case1t2}
		T_2&=&(|S|-|S_0|)\log(p\sqrt{n}) =\frac{|S|-|S_0|}{2}\log(p^2n) \nonumber \\
		&\ge& -\frac{|S_0|}{2}\log(p^2n)
	\end{eqnarray}
	By (\ref{case1t1}) and (\ref{case1t2}), 
	\begin{align*}
		T_1 + T_2 &\ge \frac{\delta(S)(1+o_p(1))}{2} - \frac{|S_0|}{2}\log(p^2n)\\
		&=\frac{\log n}{2}\left\{\frac{\delta(S)(1+o_p(1))}{\log n}-\frac{\log(p^2n)}{\log n}\right\}
	\end{align*}
	
	Since $p = O(n^\gamma)$ as $n\to\infty$ for some fixed $\gamma$, which means
	$$-\frac{\log(p^2n)}{\log n} = O(1).$$
	Thus, we have
	$$\min_{S_0\not\subset S}\text{MDL}(S)-\text{MDL}(S_0) \rightarrow \infty.$$

	\noindent{\bf{Case 2}: $S_0\subset S$.}
	
	Let $\mathcal{S}^*=\{S: S\in\mathcal{S}, S_0\subset S, S\neq S_0\}$ and $\mathcal{S}^*_j=\{S: |S|=j, S\in \mathcal{S}^*\}$.
	
	When $S_0\subset S$, we have $(\bm{I}-\bm{P}_S)\bm{\mu}=(\bm{I}-\bm{P}_S)\bm{X}_{S_0}\bm{\beta}_{S_0}=0$. Therefore, $\bm{y}^T(\bm{I}-\bm{P}_S)\bm{y}=\bm{\varepsilon}^T(\bm{I}-\bm{P}_S)\bm{\varepsilon}$.  Also
	\begin{align*}
		\text{RSS}_{S_0}-\text{RSS}_{S} & =\bm{\varepsilon}^T(\bm{I}-\bm{P}_{S_0})\bm{\varepsilon}-\bm{\varepsilon}^T(\bm{I}-\bm{P}_S)\bm{\varepsilon} \\
		& =\bm{\varepsilon}^T(\bm{P}_S-\bm{P}_{S_0})\bm{\varepsilon} \\
		& =\chi_{|S|-|S_0|}^2(S),
	\end{align*}
	where $\chi_{|S|-|S_0|}^2(S)$ follows chi-square distribution with degrees of freedom $|S|-|S_0|$.
	
	Let $c_j=2j\{\log p+\log (j\log p)\}$. According to Lemma \ref{chisq}, when $1\leq j\leq k|S_0|-|S_0|$, we have
	\begin{align*}
		P(\max_{S\in\mathcal{S}^*_j} \chi_j^2(S) \geq c_j)&\le\sum_{i=1}^{j}P(\max_{S\in\mathcal{S}^*_i} \chi_i^2(S) \geq c_j) \\ &\le\sum_{i=1}^{j}\binom{p-|S_0|}{i} P( \chi_i^2(S) \geq c_j) \\
		&\le\sum_{i=1}^{j}\binom{p}{i} P( \chi_i^2(S) \geq c_i) \rightarrow 0.
	\end{align*}
	Therefore, $\chi_{|S|-|S_0|}^2(S)\leq c_{|S|-|S_0|}(1+o_p (1))$ and
	\begin{align*}
		T_1&=\frac{n}{2}\log\left(\frac{\text{RSS}_S}{\text{RSS}_{S_0}}\right) \\
		&=-\frac{n}{2}\log\left\{1+\frac{\chi_{|S|-|S_0|}^2(S)}{\text{RSS}_{S_0}-\chi_{|S|-|S_0|}^2(S)}\right\}\\
		&\geq -\frac{n}{2}\left\{\frac{\chi_{|S|-|S_0|}^2(S)}{\text{RSS}_{S_0}-\chi_{|S|-|S_0|}^2(S)}\right\}.
	\end{align*}
	Since $n^{-1}\text{RSS}_{S_0}\rightarrow \sigma^2 = 1$ as $n\rightarrow \infty$, we have $\text{RSS}_{S_0}=n(1+o(1))$. Note that $c_{|S|-|S_0|} = o(n)$,
	\begin{eqnarray} 
		\label{case2t1}
		T_1&\geq& -\frac{c_{|S|-|S_0|}}{2}(1+o_p(1)) \nonumber \\
		&=& -(|S|-|S_0|)\log p(1+o_p(1))
	\end{eqnarray}
	
	uniformly over $\mathcal{S^*}$, and
	\begin{align} \label{case2t2}
		T_2=\frac{|S|-|S_0|}{2}\log(p^2n). 
	\end{align}
	
	For case $2$, by (\ref{case2t1}) and (\ref{case2t2}), 
	\begin{eqnarray*}
		T_1+T_2
		&\geq& \frac{(|S|-|S_0|)\log n}{2}\biggl\{\frac{\log (p^2n)}{\log n} \\
		& &-\frac{2\log p}{\log n}(1+o_p(1)) \biggl\} \\
		&= &\frac{(|S|-|S_0|)\log n}{2}\rightarrow \infty.
	\end{eqnarray*}
	Therefore, we have
	$$\min_{S_0\subset S}\text{MDL}(S)-\text{MDL}(S_0) \rightarrow \infty.$$
	
	Combing case $1$ and case $2$, it completes the proof for Theorem~\ref{thm:1}.
	
\end{proof}


\end{document}